\documentclass[acus]{JAC2003}

\usepackage{graphicx}
\usepackage{booktabs}
\usepackage{subfigure}

\usepackage{bm}% bold math

\usepackage{verbatim} % this is needed for \begin{comment} ... \end{comment}
\usepackage{lscape} % this is needed for occasional landscape figures

\usepackage{graphicx}

\usepackage{amsmath}
\usepackage{amsfonts}
\usepackage{amssymb}

\usepackage{relsize}

\usepackage{appendix}
\usepackage{subfigure}
\usepackage{colortbl}
\usepackage{empheq}
\usepackage{framed}
\usepackage{comment}

\usepackage{datetime}

\begin{document}
\title{RELATIVISTIC STERN-GERLACH DEFLECTION}
\author{Richard Talman, 
Laboratory for Elementary-Particle Physics,
Cornell University
}

\maketitle

\section{Abstract}
Modern advances in polarized beam control should make it possible
to accurately measure Stern-Gerlach (S-G) deflection of relativistic beams. 
Toward this end a relativistically covariant S-G formalism is developed 
that respects the opposite behavior under inversion of electric and magnetic 
fields. Not at all radical, or even new, this introduces a distinction
between electric and magnetic fields that is not otherwise present in 
pure Maxwell theory. Experimental configurations (mainly using polarized
electron beams passing through magnetic or electric quadrupoles) are 
described. Electron beam preparation and experimental methods needed 
to detect the extremely small deflections are discussed.

\section{Introduction}
Our purpose is to produce a relativistically-valid 
(though not quantum mechanical) theory of the deflection of a
particle, such as an electron (or more practically, a beam of
electrons), in a non-uniform electromagnetic magnetic
field. We refer to such deflection as ``Stern-Gerlach (S-G) 
deflection'', but without intending to imply that quantum-mechanical 
effects, such as the splitting of quantized states, can be
enabled.

We also neglect spin precession. This will limit the applicability
of the formulation to the passage through magnets short enough for
spin precession (calculable by the BMT equation) to be negligible.
Furthermore, for simplicity, we initially consider only on-axis 
passage through a quadrupole magnet. As it happens, since the
magnetic field on the quadrupole axis vanishes, in this case there 
will be neither electromagnetic deflection nor spin precession. 
But there will be S-G deflection, leading to (still negligible)
electromagnetic effects in subsequent approximation.

Physical constants to be used in this paper for electron charge
and electron magnetic moment are
\begin{align}
e &= 1.60217662\times10^{-19}\,{\rm C}, \notag\\
e_e^* &= -e, \notag\\
\mu_B  &= \frac{e\hbar}{2m_e} = 5.7883818066\times10^{-5}\,{\rm eV/T}, \notag\\
\mu^*_e &= (−2.00231930436182/2)\,\mu_B.
\label{eq:Intro.1}
\end{align}
These values are given to such exaggerated accuracy only to emphasize that they
are experimentally-measured and well known. However a copying error on one of 
these values, say in the fifth decimal place, would have no practical effect 
whatsoever on the experimentally challenging deflections predicted in this paper 
(because they are so small). Furthermore there will be no discussion of
subtleties such as anomalous magnetic moments. 

For reference, the fine structure constant and the speed of light
are given by
\begin{align}
\alpha 
 &=
\frac{e^2}{4\pi\epsilon_0\hbar c}
  =
\frac{1}{137.03600}, \notag\\
c &= 2.99792458\times10^8\,{\rm m/s}.
\label{eq:Intro.3m}
\end{align}
One slightly off-key note in constants~(\ref{eq:Intro.1}) may 
be noticeable; $e_e^*$ and $\mu^*_e$ have astersisks attached, suggesting
that, under some circumstances, for example in different frames of reference,
their values might be different. For the case of $e_e$ we know that charge
is a true scalar and that this will never happen, so introducing the symbol
$e_e^*$ is pure pedantry; following tradition, the asterisk will therefore
be dropped for electric charge. 
But, multiplying a vector, the magnetic moment $\mu^*_e$
is more subtle, and examples of $\mu_e$ depending on reference frame
exist in the literature. Following Conte\cite{Conte-Stern-Gerlach} the 
asterisk on $\mu^*_e$ will be adopted. The interpretation of the asterisk
(which, for consistency, would also be attached to charge $e$) is that
$\mu^*_e$ must never have any value other than that given in 
Eqs.~(\ref{eq:Intro.1}).

Other off-key steps will also be taken, the most troubling of which may
be the (temporary) divorce of electric and magnetic fields. 
Deviating from Coulomb, it was Faraday who visualized electric field lines 
emerging
from point charges. This made it natural, for example to Poisson, to visualize
magnetic field lines emerging from north or south magnetic poles. But
Oersted and Amp\'ere made it more natural to treat currents as the
sources of magnetic field. And Maxwell made it all but axiomatic to treat
electric and magnetic fields as so married as to be inseparable. 
Here (for purely pedagogical purposes) we take the (temporary) retrograde step of
divorcing electric and magnetic fields (while being careful not
to contradict well understood phenomena). 

Just as it is electric field ${\bf E}$ that applies a force to a charge
$e$ at rest, it will be (non-uniform) magnetic field ${\bf B}$
that applies a force to magnetic moment $\mu_e^*$ at rest. In this
paper we consider only electric and magnetic fields that are
transverse to particle motion. (With just one peripheral exception) 
we consider only (almost) straight line motion along a ``longitudinal'' 
axis. Lorentz boosts occur only along this axis, and fields that are 
transverse in any frame are transverse in every frame. 
The vector operator $\nabla$ can therefore, if one
wishes, be everywhere replaced by $\nabla_{\perp}$. (This would not
be valid at entrances and exits of magnets or electric elements, but
we are neglecting such end effects, if necessary assuming they
cancel in pairs.) Other than time-variation associated with entering
or exiting deflection elements, all fields are assumed to be 
time-independent. 

Expressed as equations, we take, as \emph{defining} 
relations for rest frame electric and Stern-Gerlach forces, 
\begin{align}
{\bf F}^{\rm elec} &= -e{\bf E}, \label{eq:Intro.2a}\\
{\bf F}^{SG} &= \frac{\mu_e^*}{c}\,({\bf\hat s}^*\cdot\pmb\nabla){c\bf B}\notag\\
            &= \mu_e^*\,\pmb\nabla({\bf\hat s}^*\cdot{\bf B}).
\label{eq:Intro.2b}
\end{align}
Here, ${\bf\hat s}^*$ is a unit spatial 3-vector specifying the 
orientation of the spin angular momentum in the rest frame. Only short
elements, for which ${\bf\hat s}^*$ can be treated as constant, will be
considered. In this case, since $\nabla\times{\bf B}=0$, it is valid to move
${\bf\hat s}^*$ inside the $\nabla$ operator in the second of 
Eqs.~(\ref{eq:Intro.2b}).
This has promoted magnetic moment (over magnetic charge) 
to rank with electric charge as the partners of magnetic and electric fields 
respectively. Factors of $c$ introduced, then cancelled, 
in Eq.~(\ref{eq:Intro.2b})
are an artifact of MKS units, the result of $E$ and $cB$ having the same 
units. To the extent that it is ``natural'' for the magnitudes of $E$ 
and $cB$ to be comparable, the ratio of Stern-Gerlach to electromagnetic force is
determined by a ratio of coupling constants:
\begin{equation}
\frac{\mu_B/c}{e} 
 =
1.930796\times 10^{-13}\,{\rm m},
\label{eq:Intro.3}
\end{equation}
where, except for anomalous magnetic moment and sign, Bohr
magneton $\mu_B$ is the electron magnetic moment. This
ratio has the dimension of length, compensating the inverse
length coming from the spatial derivative in Eq.~(\ref{eq:Intro.2b}).

\section{Brief Historical Perspective}
It seems fair to say that the role played by the Stern-Gerlach force in the
development of quantum mechanics has been almost as important as electric
field has been in the development of electromagnetic theory. It is curious, then, 
that Eq.~(\ref{eq:Intro.2a}) has been confirmed to accuracy approaching
that implied by physical constants~(\ref{eq:Intro.1}), while 
Eq.~(\ref{eq:Intro.2b}) has never been confirmed to much better than the 10
percent or so accuracy of the original Stern-Gerlach 
experiment\cite{Friedrich-Herschbach}\cite{Porter-SG}. 

The poor quality (even after a century) of experimental checks of 
Stern-Gerlach deflection can probably be ascribed to the smallness of 
ratio~(\ref{eq:Intro.3}). The present paper is motivated partly by the desire to 
improve this experimental determination. Since such a check is likely to
use a particle storage ring, it is essential to generalize
Eq.~(\ref{eq:Intro.2b}) to moving particles, especially having speed $v$ 
approaching the speed of light $c$. 

It might be thought that these generalizations are already
well known and uncontroversial.
Certainly, the relativistic version of 
Eq.~(\ref{eq:Intro.2a}) is well known, initially by Maxwell,
later by Einstein, and thoroughly described, for example by 
Jackson\cite{Jackson}. Furthermore the precession of the
spin vector itself, in relativistic motion, has been predicted 
by Thomas and by Bargmann, Michel, and Telegdi (BMT) and confirmed 
experimentally. 

However the Stern-Gerlach deflection of relativistic electrons,
as well as being controversial theoretically, has never been observed 
experimentally. That is, the influence of a particle's spin orientation
(whose evolution is assumed known) on a relativistic particle's orbit 
is not well understood. The orbit influence is known, however, to be so small 
that a further iteration to describe any resulting perturbation of 
the spin orientation would be gratuitous.

This paper is concerned with just this single aspect of the
Stern-Gerlach phenomenon; namely spin-dependent particle 
deflection, to be referred to here as Stern-Gerlach (S-G) particle 
deflection. Unlike the original S-G experiment, this does not
need any quantum-mechanical separation of spin states in the
uniform magnetic field component of an applied magnetic field.
Rather it is assumed that a polarized beam, prepared upstream,
passes on-axis through a quadrupole representing the non-uniform 
magnetic field that causes S-G deflection.

There exists a Bohr-Pauli ``theorem'' (or at least argument) proving the
impossibility of replicating the original Stern-Gerlach experiment
with electrons. The argument is clearly explained by Mott
and Massey\cite{MottMassey}, especially in their Figure~36. 
The argument combines the inevitable transverse beam sizes implied by 
the Heisenberg uncertainty principle (for a beam with finite 
energy spread) with the extreme weakness of the on-axis S-G bending 
relative to off-axis electromagnetic bending. In modern accelerator
jargon, even with the beam height being minimized
by a beam waist at the deflection point, rotation of the phase space
beam ellipse downstream of a quadrupole overwhelms the S-G beam shift of
any single particle relative to the full beam height.

Itself always controversial, this Bohr argument, in any case, applies to the 
deflection of single electrons, for their eventual downstream separation.
It does not apply directly to the downstream centroid shift of a beam 
that has been pre-polarized upstream of the S-G deflecting magnet. 
Nowadays, with beam centroid shifts small compared to beam size being 
observed routinely (for example in Schottky detectors) this argument has 
become somewhat suspect. 

It is argued here that modern storage ring 
developments have made it possible to measure the extremely small orbit 
centroid shifts caused by the S-G deflection of a pre-polarized beam 
of electrons, thereby making Eq.~(\ref{eq:Intro.2b}) more precise.

This is not intended to include any claim that an unpolarized electron 
beam can be split into two polarized electron beam by S-G deflection.
And, in fact, it seems inevitable that any S-G orbit shift (presumeably
of order 1\AA or less at most locations in a storage ring) will always 
be less than achievable electron beam sizes.

\section{Lorentz Force Law}
The relativistic generalization of electromagnetic force to
moving particles requires another empirical law, the Lorentz force law, 
which is reviewed next, mainly to make the electromagnetic field tensor
available:
\begin{equation}
{\bf F}^{\rm e.m.} = -e({\bf E} + e{\bf v}\times{\bf B}).
\label{eq:ElecForce.1}
\end{equation}
Following Jackson, and/or Steane\cite{Steane}, but skipping many 
details, we introduce contravariant (upper index) and covariant 
(lower index) tensors, shown (in various representations) in 
Table~\ref{tbl:RelNotation}. 4D dot product invariants are defined
as by Steane, except with reversed metric tensor signs. 
This dot product notation allows many subscripts and superscripts to
be suppressed and the distinction between contravariant and 
covariant components to be largely hidden.
\begin{table*}
\caption{\label{tbl:RelNotation}The main 4-vectors to be used,
as well as the 2-index, anti-symmetric electromagnetic 4-tensor $\mathbb{F}$. Font
selection can be inferred from this table. Typically unprimed variables refer to a 
general frame of
reference, such as the laboratory, and primed variables refer to a frame in which
the particle is either at rest, or moving at non-relativistic velocity. Transverse
spin vector ${\bf s}_{\perp}$ and longitudinal spin component ${\bf s}_{\parallel}$
Lorentz transform differently, as given in Eq.~(\ref{eq:Pauli-Lubarski.4}).}
\medskip
\begin{tabular}{|c|c|c|c|c|c|c|c|}  \hline
 symbol       & definition        & contravariant components     & rest frame              &   name                 &  invariant    \\ \hline
   X          &    X              & $(ct,{\bf r})$               & $(ct',{\bf r'})$        & 4-displacement         &  $c^2\tau^2$   \\ 
   U          & $d{\rm X}/d\tau$  & $(\gamma c,\gamma{\bf v})$   & $(c,0)$                 & 4-velocity             &  $c^2$         \\
   P          & $m_e$U            & $(E/c,{\bf p})$              & $(mc,0)$                & 4-energy-momentum      & $m^2c^2$       \\
   W          &  Eq.~(\ref{eq:Pauli-Lubarski.1})                
                                  & $({\bf s}\cdot{\bf p}, (\mathcal{E}/c){\bf s})$ 
                                                                 & $(0, m_ec{\bf s'})$     & Pauli-Lubanski 4-spin  &  -$m^2c^2$   \\
   S          & S=W/$(mc)$ &$(\gamma{\bf s}\cdot\pmb\beta,\gamma{\bf s}$) & $(0,{\bf s'})$ 
                                                                                           &  4-spin        &  $-|{\bf s}'|^2$ \\
  $\Box$      &                   & $(\partial/\partial(ct), \pmb\nabla)$
                                                                 & $(\partial/\partial(c\tau), \pmb\nabla')$                     
                                                                                           &  4-gradient            &          \\
\hline
 $\mathbb{F}$ &                   & Eq.~(\ref{eq:ElecForce.2})   &                         &  EM-field-tensor       &          \\
\hline
\end{tabular}
\end{table*}
An ``electric'' (as unconventionally contrasted here with ``magnetic'') 
field tensor is defined by
\begin{equation}
\mathbb{F}({\bf E}, \cdot)
 =
\begin{pmatrix} 
  0    & E^1/c & E^2/c & E^3/c \\
-E^1/c &  0    & -B^3  & B^2  \\
-E^2/c &  B^3  &   0   & -B^1   \\
-E^3/c & -B^2  &  B^1  &   0 
\end{pmatrix}
\label{eq:ElecForce.2}
\end{equation}
(Also unconventionally) this tensor is formally expressed here as a function
$\mathbb{F}({\bf E}, \cdot)$ of the electric field \emph{vector}, 
irrespective of the fact that it
depends also on the components of ${\bf B}$ which, if present at all, are subsidiary
and are represented only by a dot in the argument list. 

This notation favors reference frames in which the magnetic field actually vanishes,
since the only non-vanishing elements are in the first row (which influences energy
changes) and the first column (which influences momentum changes). 
It will soon be clear that this notation is really just an insignificant, 
artificial way of guiding the discussion.
Having said earlier, in connection with Eqs.~(\ref{eq:Intro.2a}) 
and (\ref{eq:Intro.2b}), that ${\bf E}$ and ${\bf B}$ fields are not
to be treated as components of the same ``physical object'', it
probably seems to be cheating to combine them, as in this equation.
The $B^i$ components in this equation will shortly be associated 
with the $B^i$ components in Eq.~(\ref{eq:Intro.2b}) 
and it will be convenient then, not to have to 
change their symbols.

Eq.~(\ref{eq:ElecForce.2}) also differs slightly from Jackson's 
conventions in that the
components are expressed with raised indices, in spite of the
fact that there is no contravariant/distinction for 3D vector components,
which are usually given lower indices. When matching 3D coordinates 
with 4D coodinates, while encouraging the use, by default, of contravariant 
indices for 4D fields and coordinates, it is less confusing to use
upper indices for 3D components. 

There is a more significant issue with definition~(\ref{eq:ElecForce.2}).
As Fitzpatrick\cite{Fitzpatrick} explains, Eq.~(\ref{eq:ElecForce.2}), as
written, seems inconsistent with our understanding that ${\rm B}$ is
a pseudo-vector, while ${\rm E}$ is a vector. Reflection or transition
from right- to left-handed coordinate axes, would change the meaning
of the equation. It might seem to be less inconsistent
to express the magnetic components as $B^1=B_{23}$, $B^2=B_{31}$, 
$B^3=B_{12}$, but this has, effectively, already been accomplished 
by shifting the positions of the magnetic components in the matrix. As long 
as one uses only coordinate frames not involving reflections, it is legitimate
to mix vectors and pseudovectors in this way and, following convention, we 
tolerate this limitation. 

Applying Eq.~(\ref{eq:ElecForce.2}), Newton's law for the Lorentz force 
law can be written in abbreviated form as
\begin{equation}
m_e\,\frac{d{\rm U}}{d\tau} 
 =
-e\,\mathbb{F}({\bf E}, \cdot)\cdot{\rm U}. 
\label{eq:ElecForce.3}
\end{equation}
Expressed (as matrix product) in the rest frame, 
this produces covariant acceleration components, expressed as a column
matrix
{
\small
\begin{align}
& \begin{pmatrix} d{\rm U^0}'/d\tau & 
                 -d{\rm U^1}'/d\tau & 
                 -d{\rm U^2}'/d\tau & 
                 -d{\rm U^3}'/d\tau 
\end{pmatrix}^T 
%\frac{d{\rm U'}}{d\tau} 
 =\notag\\
&\qquad \frac{-e}{m_e}\,
\begin{pmatrix} 
    0     & {E^1}'/c & {E^2}'/c &  {E^3}'/c \\
-{E^1}'/c &   0      & -{B^3}'  &  {B^2}'   \\
-{E^2}'/c & {B^3}'   &   0      & -{B^1}'   \\
-{E^3}'/c & -{B^2}'  & {B^1}'   &   0 
\end{pmatrix}\,
\begin{pmatrix} c \\ 0 \\ 0 \\ 0 \end{pmatrix},
\label{eq:ElecForce.4}
\end{align}
}

\noindent
where the ${B^i}'$ components are ineffective not because
they vanish, but because they multiply velocity
components which do vanish. In this frame it seems legitimate
to express $\mathbb{F}$ as a function of only ${\bf E}$.
In the laboratory frame $\mathbb{F}({\bf E}, \cdot)$ is given by 
Eq.~(\ref{eq:ElecForce.2}) and U by $(\gamma c, \gamma{\bf v})^{\rm T}$.

\section{Relativistic Stern-Gerlach Deflection}
The least familiar entry in Table~\ref{tbl:RelNotation}
is the Pauli-Lubanski spin tensor W, which is 
a momentum-weighted spin vector defined by covariant components
\begin{align}
W_a &= \frac{1}{2}\,\epsilon_{\lambda a\mu\nu}{\rm P}^{\lambda}{\rm S}^{\mu\nu}\notag\\
    &= \Big((s_{\parallel}\hat{\pmb\beta} + {\bf s}_{\perp})\cdot{\bf p},\,
            -(1/c)\mathcal{E}(s_{\parallel}\hat{\pmb\beta} + {\bf s}_{\perp})
\Big)
\label{eq:Pauli-Lubarski.1}
\end{align}
where $\epsilon_{\lambda a\mu\nu}$ is a standard 4-index tensor with non-vanishing 
components equal to $\pm1$, depending on whether the indices are an
even or odd permutation of 0,1,2,3, (and hence anti-symmetric in all indices), 
and ${\rm S}^{\mu\nu}$ is a 2-index spin tensor. 

Suppose that, relative to the rest frame,
the laboratory frame is moving with velocity $-v{\bf\hat x}$. 
The inverse boost coordinate transformation giving $(ct,{\bf x})$ 
in terms of $(ct',{\bf x'})$ is
\begin{align}
ct &= \gamma(ct' + \pmb\beta\cdot{\bf x'})\notag\\
{\bf x}
 &= 
{\bf x'} + \Big(-\gamma ct' 
         + \frac{\gamma^2}{1+\gamma}\,\pmb\beta\cdot{\bf x'}\Big)\,\pmb\beta.
\label{eq:Pauli-Lubarski.2p}
\end{align}
For W to be a valid 4-vector requires it to be subject to the same 
Lorentz transformation. This requires the (covariantly-expressed)
component expressions on the right hand side of Eq.~(\ref{eq:Pauli-Lubarski.1}) 
to be valid in all frames.
To check that this is indeed the case,
one can perform Lorentz boost~(\ref{eq:Pauli-Lubarski.2p})
to 4-vector ${\rm W'}$ to produce contravariant laboratory frame components
\begin{align}
 s_{\parallel}p 
   \overset{?}{\ =\ } 
&{\rm W^0}         =  
\gamma{\pmb\beta}\cdot mc{\bf s'}
                 = \gamma mv s'_{\parallel}
                 = s'_{\parallel}p,
\label{eq:Pauli-Lubarski.3a}\\
(1/c)\mathcal{E}(s_{\parallel}\hat{\pmb\beta} &+ {\bf s}_{\perp})
       \overset{?}{\ =\ } \notag\\
{\bf W} &=  
mc{\bf s'} + \frac{\gamma^2}{1+\gamma}\,\beta^2 mcs'_{\parallel}\hat{\pmb\beta}\notag\\
                &=     mc({\bf s'} - s'_{\parallel}\hat{\pmb\beta})
                  + mc\Big(1 + \frac{\gamma^2\beta^2}{1+\gamma}\Big)\,s'_{\parallel}\hat{\pmb\beta}\notag\\
                &= (1/c)\mathcal{E}'\Big(s'_{\parallel}\hat{\pmb\beta} + \frac{{\bf s'}_{\perp}}{\gamma}\Big).
\label{eq:Pauli-Lubarski.3b}
\end{align}
where the questioned equalities on the left can be answered in
the affirmative if and only if the right hand side of 
Eq.~(\ref{eq:Pauli-Lubarski.1}) is valid in all reference frames.
To make these formulas self-consistent, and to answer the questions in the 
affirmative requires
\begin{equation}
s_{\parallel} = s'_{\parallel},
\quad\hbox{and}\quad 
{\bf s}_{\perp} = \frac{{\bf s'}_{\perp}}{\gamma}.
\label{eq:Pauli-Lubarski.4}
\end{equation}
These, therefore, are the Lorentz transformed components
of spin 4-vector S. (It is important to comment that these components 
will \emph{not} appear in the relativistically-invariant Stern-Gerlach
equation~(\ref{eq:LabTransverse.3}), whose derivation is
our immediate goal.)

The spatial 3-vector ${\bf s}$ has constant magnitude in any
particular reference frame, and is the spatial
part of the conventional (BMT) 4-spin vector.
In the rest frame 
\begin{equation}
{\rm W'} = mc{\rm S'} = (0, mc{\bf s'}).
\label{eq:Pauli-Lubarski.1p}
\end{equation}
The momentum weighting of the Pauli-Lubarski 4-vector
is removed when W is reduced to ``helicity''
\begin{equation}
s_{\parallel} = \frac{\bf s\cdot p}{p}.
\label{eq:Pauli-Lubarski.1q}
\end{equation}

A formula valid in one referenceframe can be extended to other frames using 
what I will call the ``Hagedorn Principle'', though the principle was 
undoubtedly introduced by someone before him, probably 
Einstein. Here the principle is copied verbatim from Hagedorn
(including the boxed format).

\medskip
\boxed{
\parbox{0.45\textwidth}
{If an equation given in a  particular Lorentz system can be
written in a manifestly covariant form (that is, both sides
have the same transformation property!) which in the particular 
Lorentz frame reduces to the equation given originally, this 
covariant form is the unique generalization of the equation given.}
}

\medskip
For invariant expression of the S-G deflection we need to follow the
same path as was used for electromagnetic deflection, 
but with one change, brought out by the fact that
the magnetic field is, in fact, a pseudo-vector. Jackson, in Eq.~(11.140),
defines a ``dual'', 2-index, covariant tensor
\begin{align}
\mathfrak{F}(\cdot, {\bf B})
 &=
\frac{1}{2}\,\epsilon_{\alpha\beta\gamma\delta}{\mathbb{F}}^{\gamma\delta}\notag\\
 &= 
\begin{pmatrix} 
 0   &  B^1   &  B^2   &  B^3   \\
-B^1 &   0    &  E^3/c & -E^2/c \\
-B^2 & -E^3/c &   0    &  E^1/c \\
-B^3 &  E^2/c & -E^1/c &   0 
\end{pmatrix}
\label{eq:S-G.1}
\end{align}
It differs from $\mathbb{F}$ by the replacements ${\bf E}/c\rightarrow {\bf B}$
and ${\bf B}\rightarrow -{\bf E}/c$. 
This time we have (artificially) expressed 
$\mathfrak{F}(\cdot, {\bf B})$ as a function only of ${\bf B}$ as a reminder that it 
is appropriate for representing pseudo-vectors. This notation is most convenient
for magnets at rest. Note that, even though $E$ and $B$ have different 
dimensions in MKS units, that ``dual'' tensors $\mathbb{F}({\bf E})$ and
$\mathfrak{F}(\cdot, {\bf B})$, in MKS units are both measured in Tesla
magnetic field units.

With S-G force given by Eq.~(\ref{eq:Intro.2b}),
copying from Eq.~(\ref{eq:ElecForce.3}), but using $\mathfrak{F}(\cdot, {\bf B})$ 
because the argument is a pseudo-vector, we obtain
\begin{equation}
m_e\,\frac{d{\rm U}'}{d\tau} 
 =
\mu_e^*\,
\mathfrak{F}\big(\cdot, ({\bf s}'\cdot\pmb\nabla'){\bf B}'\big)\cdot\frac{{\rm U}'}{c},
\label{eq:S-G.1p}
\end{equation}
as the rest frame equation of motion. The final factor is divided by $c$ 
because of MKS units. The Hagedorn principle can not
yet be applied to this formula because the argument
of $\mathfrak{F}$ is expressed as a 3-vector rather than a 4-vector.
Fortunately we are dealing with fields that are constant in time. This is
violated only at particle entry and exit from the element causing
the deflections. Even if there is a non-zero deflection on
entry, it will be cancelled on exit. In the rest frame,
where $\mathcal{E}=mc^2$ we can therefore make the replacement
\begin{equation}
{\bf s}'\cdot\pmb\nabla' = - \frac{\rm W'}{mc}\cdot\Box'
\label{eq:S-G.2}
\end{equation}
where $\Box$ is defined in Table~\ref{tbl:RelNotation}, 
to produce 
\begin{equation}
m_e\,\frac{d{\rm U}'}{d\tau} 
 =
-\mu_e^*\,\mathfrak{F}\bigg(\cdot, \Big(\frac{\rm W'}{mc}\cdot\Box'\Big)
                                   {\bf B}'\bigg)\cdot\frac{{\rm U}'}{c}.
\label{eq:S-G.3}
\end{equation}
This is now expressed in terms of invariant quantities.

Fortunately we do not need the full generality indicated by 
Eq.~(\ref{eq:S-G.3}). We are interested in the perturbation, to
a straight line on-axis orbit through a quadrupole, caused by the
S-G force. Let us say the axis in question carries the label ``1''.
(For standard accelerator coordinates ``1'' would naturally be
replaced by ``$z$''. But standard discussions of Lorentz boosts
usually assume a particle is moving along the ``$x$'' axis in
the laboratory, which is why we make this choice.)
We are also restricting the generality by assuming that the longitudinal 
components of both ${\bf E}$ and ${\bf B}$ vanish in both the laboratory 
and rest frames. 

For simplicity we can also treat purely transverse
and purely longitudinal polarization separately. Let us start with
just transverse polarization; i.e. replace ${\bf s}$
by ${\bf s}_{\perp}$, meaning (for motion along the $x^1$-axis)
that $s_1=0$. With these specializations the 
rest frame equation of motion becomes
\begin{equation}
m_e\,\frac{d{\rm U'}}{d\tau} 
 =
-\mu_e^*\,\mathfrak{F}\bigg(\cdot, \Big(\frac{\rm W'_{\perp}}{mc}\cdot\Box'_{\perp}\Big){\bf B'_{\perp}}\bigg)\cdot\frac{{\rm U}'}{c}.
\label{eq:S-G.4}
\end{equation}
Under our assumed special conditions, the transverse $(x^2,x^3)$ coordinates 
are conserved in the Lorentz transformation along the $x^1$ axis. These 
coordinates can therefore be treated as parameters, immune from alteration
during boosts along the particle velocity axis. 
Also, because of the linearity of the matrix multiplication,
the expression in large parentheses can be moved from argument to coefficient;
\begin{equation}
m_e\,\frac{d{\rm U'}}{d\tau} 
 =
-\mu_e^*\,\Big(\frac{\rm W'_{\perp}}{mc}\cdot\Box'_{\perp}\Big)
         \mathfrak{F}\big(\cdot, {\bf B'_{\perp}}\big)\cdot\frac{{\rm U}'}{c}.
\label{eq:S-G.4p}
\end{equation}
Applying the Hagedorn principle, the laboratory frame equation of motion is
\begin{equation}
m_e\,\frac{d{\rm U}}{d\tau} 
 =
-\mu_e^*\,\Big(\frac{\rm W_{\perp}}{mc}\cdot\Box_{\perp}\Big)
         \mathfrak{F}\big(\cdot, {\bf B_{\perp}}\big)\cdot\frac{{\rm U}}{c}.
\label{eq:S-G.4q}
\end{equation}
After these manipulations, Eq.~(\ref{eq:S-G.4}) can more readily be compared
with the electromagnetic force equation~(\ref{eq:ElecForce.3}). In particular,
$\mathfrak{F}\big(\cdot, {\bf B_{\perp}}\big)\cdot{\rm U}/c$ is relativistically 
invariant, like $\mathbb{F}({\bf E}, \cdot)\cdot{\rm U}$. And the operations
acting on $\mathfrak{F}\big({\bf B_{\perp}}\big)\cdot{\rm U}/c$ in 
Eq.~(\ref{eq:S-G.4}) preserve this frame invariance.

According to Eq.~(\ref{eq:Pauli-Lubarski.4}) rest frame transverse
polarization is also transverse in the laboratory.
And the azimuthal polarization angle is the same in
rest frame and lab frame. Expressed in terms of the rest 
frame polarization vector ${\bf s}'_{\perp}$, the laboratory frame polarization 
vector is 
\begin{equation}
{\bf s}_{\perp} = \frac{1}{\gamma}\,{\bf s}'_{\perp}
               = \frac{{s^2_x}'{\bf\hat x_2} +  {s^3_x}'{\bf\hat x_3}}{\gamma}.
\label{eq:LabTransverse.1}
\end{equation}
The reason this replacement is appropriate is that the 
transverse magnetic moment vector 
is defined (only in the rest frame) to be $\mu_e^*{\bf s'}_{\perp}$. 
For substitution into laboratory equation of motion~(\ref{eq:S-G.4}),
and including longitudinal polarization terms as well, we have
\begin{align}
\frac{W_{\perp}}{m_ec} 
  &= 
\bigg(
0, \frac{\mathcal{E}}{m_ec^2}\frac{{s^2_x}'{\bf\hat x_2} +  {s^3_x}'{\bf\hat x_3}}{\gamma}
\bigg)                                                           \notag\\
  &= (0,\ {s^2_x}'{\bf\hat x_2} +  {s^3_x}'{\bf\hat x_3}), \notag\\
\Box_{\perp} 
 &= 
\Big( 0,\ 
   {\bf\hat x_2}\,\frac{\partial}{\partial x^2} + {\bf\hat x_3}\,\frac{\partial}{\partial x^3}  
\Big), \notag\\
\frac{W_{\parallel}}{m_ec} 
  &= 
\bigg(
\gamma\,\frac{v}{c}, 0 
\bigg),\quad
\Box_{\parallel} = (0,{\bf\hat x_1}\partial/\partial x^1),\notag\\
\frac{W}{m_ec}\cdot\Box 
&= 
-{s^2_x}'\,\frac{\partial}{\partial x^2} - {s^3_x}'\,\frac{\partial}{\partial x^3}
=
-{\bf\hat s}^*\cdot\pmb\nabla_{\perp}.
\label{eq:LabTransverse.2}
\end{align}
In the final step we have taken advantage of ${\rm W}_{\parallel}\cdot\Box_{\parallel}=0$,
and that all fields are independent of $x^1$. Also
we have re-introduced the same rest frame unit vector ${\bf\hat s}^*$
as appeared in Eq.~(\ref{eq:Intro.2b}). Its asterisk emphasizes that it is a unit
vector in the rest frame, in spite of the fact that all other quantities in the 
equation of motion, including $\pmb\nabla_{\perp}$, refer to the 
laboratory frame. 

Finally, substituting from Eqs.~(\ref{eq:LabTransverse.2}) into Eq.~(\ref{eq:S-G.4}),
for transverse fields independent of time and longitudinal position, the frame-invariant
Stern-Gerlach equation of motion is
\begin{equation}
m_e\,\frac{d{\rm U}}{d\tau} 
 =
\mu_e^* {\bf\hat s}^*\pmb\cdot\,
    \mathfrak{F}\big(\cdot,\pmb\nabla{\bf B}\big)\cdot\frac{{\rm U}}{c}.
\label{eq:LabTransverse.3}
\end{equation}
This formula is especially convenient in reference frames in which ${\bf E}$=0.
As before, in reference frames where the electric field does not, in fact,
vanish, the $\cdot$ in the first argument has to be replaced by $\pmb\nabla{\bf E}$.

\section{Deflection Examples}
Transverse fields (relative to their velocity ${\bf v}$,
which we will take to be the ${\bf\hat x_1}$-axis) in the laboratory 
(unprimed) and in the electron rest frame (primed), are 
related by
\begin{align}
{\bf E}' &= \gamma({\bf E} + {\bf v}\times {\bf B}),\notag\\
{\bf B}' &= \gamma({\bf B} - {\bf v}\times {\bf E}/c^2).
\label{eq:LorentzFieldTrans.1}
\end{align}
Expressed in terms of components,
\begin{align}
{E^2}' &= \gamma(E^2-vB^3),\quad {E^3}' = \gamma(E^3+vB^2), \notag\\
{B^2}' &= \gamma(B^2+vE^3/c^2),\quad {B^3}' = \gamma(B^3-vE^2).
\label{eq:LorentzFieldTrans.2}
\end{align}

\subsection{Particle Deflection in Electrostatic Separator}
For practice with the covariant formulation, using entries from Table~\ref{tbl:RelNotation}, 
we consider the deflection of an electron of velocity $v{\bf\hat x}_1$ 
in the transverse electric field $E{\bf\hat x_2}$ in a laboratory electrostatic
separator of length $L_E$. Using $d/d\tau=\gamma\,d/dt$, Eq.~(\ref{eq:ElecForce.3}),
in laboratory coordinates, 
produces contravariant time derivative
\begin{align}
-\frac{dp^2}{dt}   % why no minus sign because LHS is covariant?
 =
\frac{-e}{\gamma}\,
\begin{pmatrix} 
   0  & 0 &  E/c & 0 \\
   0  & 0 &   0  & 0 \\
 -E/c & 0 &   0  & 0 \\
   0  & 0 &   0  & 0 
\end{pmatrix}\,
\begin{pmatrix} \gamma c \\ \gamma v \\ 0 \\ 0 \end{pmatrix} \Bigg|_2
 =
eE.
\label{eq:Example.2}
\end{align}
So the force is $dp^2/dt=-eE$, the duration $L_E/v$, the transverse impulse
$-eE L_E/v$, and the angular deflection is
\begin{equation}
\Delta\theta_2 \approx \frac{-eEL_E/v}{p},
\label{eq:Example.3}
\end{equation}
which is the transverse momentum impulse divided by the
longitudinal momentum. 

As a check, the same result can be obtained 
by realizing the rest frame magnetic field, though non-zero,
applies no force to a particle at rest; the rest frame 
electric field is $\gamma E{\bf\hat x_2}$; and
the separator length is foreshortened to $L_E/\gamma$,
so the electric field is present for time duration $L_E/v$.
This yields rest frame transverse momentum impulse 
$-eEL_E/v$, which is the same as the laboratory frame
transverse momentum impulse.

\subsection{S-G Deflection in Magnetic Quadrupole}
\begin{figure}[h]
\centering
\includegraphics[scale=0.35]{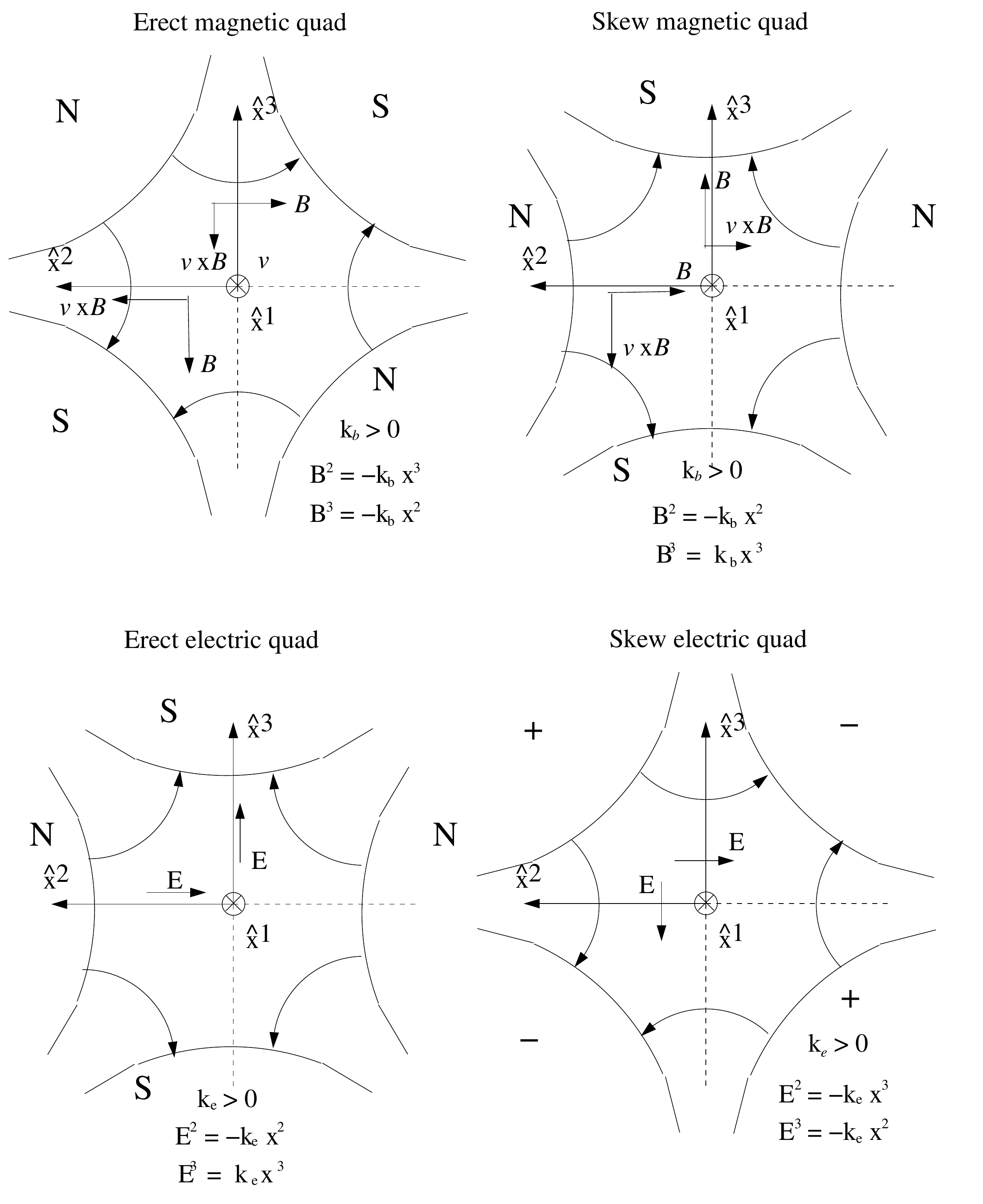}
\vskip -0.2cm
\caption{\label{fig:E-B-Quadrupoles}Fields of magnetic and electric
quadrupoles, both erect and skew. The hyperbolic curves represent
the surfaces of iron poles in the magnetic case, and electrodes in
the electric case. In both cases the field lines are normal to these
2D surfaces.
}
\end{figure}

Because their fields vanish on-axis, quadrupoles, either magnetic
or electric, erect or skew, provide ideal tests of Stern-Gerlach 
deflection. Their field boundaries and fields are shown
in Fig.~\ref{fig:E-B-Quadrupoles}. Formulas for their
field components are given in the figure.
Consider a transversely polarized electron passing on-axis through
such a magnetic quadrupole. The deflection equation is  
Eq.~(\ref{eq:LabTransverse.3}).
Cancelling $(-\gamma)$ from both sides,
\begin{align}
&\begin{pmatrix} -dp_0/dt \\ dp^1/dt \\ dp^2/dt \\ dp^3/dt \end{pmatrix}
  =  \notag\\
&\mu_e^*\,
 \begin{pmatrix} 
  0  & 0 & \cdot 
         & \cdot \\
  0  & 0 &  0  &  0  \\
\hat{s}^{*2}\,B^2_{,2} + \hat{s}^{*3}\,B^2_{,3} & 0 &  0  &  0  \\
\hat{s}^{*2}\,B^3_{,2} + \hat{s}^{*3}\,B^3_{,3} & 0 &  0  &  0 
\end{pmatrix}
\begin{pmatrix} 1 \\ v/c \\ 0 \\ 0 \end{pmatrix}.
\label{eq:relSG.5}
\end{align}

\subsubsection{Skew Magnetic Quadrupole. }
Representing partial derivatives with
``,$\mu$'' notation, field components for a \emph{skew} magnetic 
quadrupole are given by
\begin{equation}
B^2=-k_bx^2,\  B^3=k_bx^3,
\ \hbox{with}\ 
k_b=-B^2_{,2}=B^3_{,3}.
\label{eq:relSG.4}
\end{equation}
(Reminder: superscripts are indices, not powers.) Because the
magnetic field is transverse, there is no coupling to longitudinal
spin component $\hat s^{*1}$. The operative equations are
\begin{equation}
\begin{pmatrix} dp^2/dt \\ dp^3/dt \end{pmatrix}
  =  
\mu_e^*
 \begin{pmatrix} 
-\hat{s}^{*2}\,k_b & 0 \\
 \hat{s}^{*3}\,k_b & 0 
\end{pmatrix}
\begin{pmatrix} 1 \\ v/c \end{pmatrix}
 =
\mu_e^*k_b
 \begin{pmatrix} 
-\hat{s}^{*2} \\
 \hat{s}^{*3}  
\end{pmatrix},
\label{eq:relSG.6}
\end{equation}
and the momentum impulses are
\begin{align}
\Delta p^2 = -\mu_e^*\hat{s}^{*2}k_b\,\frac{L_q}{v},\quad
\Delta p^3 = \mu_e^*\hat{s}^{*3}k_b\,\frac{L_q}{v} .
\label{eq:LabTransverse.4}
\end{align}
Just as the relativistic generalization of electromagnetic deflection
was checked in the previous example, the relativistic S-G
formulation can be checked by direct Lorentz transformation of
the recoil momentum, as calculated in the rest frame. 
For example, taking $\hat{s}^{*2}$=1, $\hat{s}^{*3}$=0,
The  $\Delta {p^2}'$ momentum recoil component can be obtained 
from Eq.~(\ref{eq:Intro.2b}). Using the transformation
properties of $B^2$ and $x^2$, the rest frame S-G force is 
$\mu_e^*\gamma\partial {B^2}/\partial {x^2}{\bf\hat x_2}$.
Because the quadrupole length is foreshortened to $L_q/\gamma$,
the interaction duration is $L_q/v$. Since the recoil is transverse, 
the laboratory and rest frame momentum recoils are the same;
\begin{equation}
\Delta p^2 = -\mu_e^*k_b\,\frac{L_q}{v},
\label{eq:LabTransverse.5}
\end{equation}
in agreement with Eq.~(\ref{eq:LabTransverse.4}).

\subsubsection{Erect Magnetic Quadrupole}
Field components for an \emph{erect} magnetic quadrupole
are given by
\begin{equation}
B^2=-k_bx^3,\  B^3=-k_bx^2,
\ \hbox{with}\ 
k_b=-B^2_{,3}=-B^3_{,2}.
\label{eq:relSG.4em}
\end{equation}
The operative equations are
\begin{equation}
\begin{pmatrix} dp^2/dt \\ dp^3/dt \end{pmatrix}
  =  
\mu_e^*
 \begin{pmatrix} 
 -\hat{s}^{*3}\,k_b & 0 \\
 -\hat{s}^{*2}\,k_b & 0 
\end{pmatrix}
\begin{pmatrix} 1 \\ v/c \end{pmatrix}
 =
\mu_e^*k_b
 \begin{pmatrix} 
-\hat{s}^{*3} \\
-\hat{s}^{*2}  
\end{pmatrix},
\label{eq:relSG.6em}
\end{equation}
and the momentum impulses are
\begin{align}
\Delta p^2 = -\mu_e^*\hat{s}^{*3}k_b\,\frac{L_q}{v},\quad
\Delta p^3 = -\mu_e^*\hat{s}^{*2}k_b\,\frac{L_q}{v} .
\label{eq:LabTransverse.4em}
\end{align}
Realizing that the skew and erect magnets are, in reality,
identical, though rotated by $45^{\circ}$, on could have
obtained Eqs.~(\ref{eq:LabTransverse.4em}) by simple 
rotation of one or the other of the top figures
in Figure~\ref{fig:E-B-Quadrupoles}. The formalism
is therefore consistent, at least to this extent,
for magnetic quadrupoles.

\subsection{S-G Deflection in Electric Quadrupole}
Considering that the calculation can be patterned after
the previous examples, for laboratory frame evaluation
of the S-G deflection in an electric quadrupole,
how can we know whether to use 
$\mathbb{F}({\bf E}, \cdot)$ from Eq.~(\ref{eq:ElecForce.2})
or $\mathfrak{F}(\cdot, {\bf B})$ from Eq.~(\ref{eq:S-G.1})
as the operative laboratory electromagnetic matrix?
Apart from their different predicted recoil directions, 
especially for $v<<c$
they give greatly different recoil magnitudes. 
Even though all components of ${\bf B}$ vanish in the laboratory,
as already explained, it is $\mathfrak{F}(\cdot, {\bf B})$ that 
has the correct inversion symmetry. Again canceling $(-\gamma)$ 
from both sides, the equations of motion are 
\begin{align}
&\begin{pmatrix} -dp_0/dt \\ dp^1/dt \\ dp^2/dt \\ dp^3/dt \end{pmatrix}
  =  \notag\\
&\mu_e^*\,
\begin{pmatrix} 
 0   &   0    & \cdot  & \cdot \\
 0   &   0    & \cdot  & \cdot \\
 0   &  \hat s^{*2}E^3_{,2}/c + \hat s^{*3}E^3_{,3}/c &   0    &   0   \\
 0   & -\hat s^{*2}E^2_{,2}/c - \hat s^{*3}E^2_{,3}/c &   0    &   0 
\end{pmatrix}
\begin{pmatrix} 1 \\ v/c \\ 0 \\ 0 \end{pmatrix}.
\label{eq:relSG.5E}
\end{align}
\subsubsection{Erect Electric Quadrupole}
For an \emph{erect} electric quadrupole the field components are 
\begin{equation}
E^2=-k_ex^2,\  E^3=k_ex^3,
\ \hbox{with}\ 
k_e=-E^2_{,2}=E^3_{,3}.
\label{eq:relSG.6Ee}
\end{equation}
The operative equations are
{\small
\begin{equation}
\begin{pmatrix} dp^2/dt \\ dp^3/dt \end{pmatrix}
  =  
\mu_e^*
 \begin{pmatrix} 
  \hat s^{*3}E^3_{,3}/c & 0 \\
 -\hat s^{*2}E^2_{,2}/c & 0 
\end{pmatrix}
\begin{pmatrix} 1 \\ v/c \end{pmatrix}
 =
\frac{\mu_e^*k_ev}{c^2}
 \begin{pmatrix} 
 \hat{s}^{*3} \\
 \hat{s}^{*2}  
\end{pmatrix}.
\label{eq:relSG.7Ee}
\end{equation}
}

\subsubsection{Skew Electric Quadrupole}
For a \emph{skew} electric quadrupole the field components are 
\begin{equation}
E^2=-k_ex^3,\  E^3=-k_ex^2,
\ \hbox{with}\ 
k_e=-E^2_{,3}=-E^3_{,2}.
\label{eq:relSG.6Se}
\end{equation}
The operative equations are
{\small
\begin{equation}
\begin{pmatrix} dp^2/dt \\ dp^3/dt \end{pmatrix}
  =  
\mu_e^*
 \begin{pmatrix} 
0 &  \hat s^{*2}E^3_{,2}/c \\
0 & -\hat s^{*3}E^2_{,3}/c
\end{pmatrix}
\begin{pmatrix} 1 \\ v/c \end{pmatrix}
 =
\frac{\mu_e^*k_ev}{c^2}
 \begin{pmatrix} 
-\hat{s}^{*2} \\
 \hat{s}^{*3} 
\end{pmatrix}.
\label{eq:relSG.7Se}
\end{equation}
}

\subsection{Spin-Orbit Coupling}
\subsection{Spin-Orbit Central Force}
Stern-Gerlach deflection plays a significant role in the theory
of an atom at rest. This is an example in which the 
only laboratory field is purely electric. There is, however, magnetic
field in the rest frame of an electron (treated classically) circulating
around a point charge. Furthermore, there is non-zero radial electric
field on the electron's elliptical or, for simplicity, let us say, 
circular, orbit.

There are several reasons why the present formalism cannot be applied
directly to atomic physics. The most important, of course, is that the
analysis has been classical (though relativistic rather than Newtonian). 
At most the formalism can therefore be applied to 
``semi-classical atomic theory''.  But, since the analysis has been
stubornly Newtonian, rather than Hamiltonian, even semi-classical application
is not automatic.  Furthermore, S-G bending elemants have been
assumed short enough that the spin direction can be assumed constant
during any S-G deflection. This is certainly not valid for an atomic
orbit.

In spite of these reservations, it may have mnemonic value, in 
contemplating modern investigation of relativistic Stern-Gerlach
deflection, to pretend to apply the same formalism to atomic
orbits treated by classical mechanics.  For simplest comparison
we will consider a hydrogen atom $Z$=1, in the lowest
Bohr model semi-classical case, having $n=1$. We shall treat
this system as a tiny storage ring.
In this state the (non-relativistic) total energy is 
given by\cite{Leighton}
\begin{equation}
\mathcal{E}_1 = \frac{m_ee^4}{32\pi^2\epsilon_0^2\hbar^2(1+m_e/m_p)}.
\label{eq:Bohr.1}
\end{equation}
and the orbit radius by
\begin{equation}
a_1 = \frac{4\pi\epsilon_0\hbar^2}{m_ee^2}
    = 5.29177\times10^{-11}\,{\rm m}.
\label{eq:Bohr.2}
\end{equation}

(In our coordinate convention with indices as superscripts) 
the particle orbit radius is $r=a_1-x^2$ where $x^2$ is a 
radially-inward displacement, meaning the unit vector ${\bf\hat x}_2$ 
always points from the electron toward the nucleus proton.
The potential energy, and the radial electric field are
\begin{align}
V(r) &= -\frac{e^2}{4\pi\epsilon_0 r}, \label{eq:Bohr.3}\\
E^2(r) &= -\frac{e}{4\pi\epsilon_0 (a_1-x^2)^2}
        = -E_1\,\Big(1 + \frac{2x^2}{a_1}+\cdots\Big).
\notag
\end{align}
where $E_1$ is positive by definition, and its ``1'' subscript
indicates ``first'' Bohr orbit. For Stern-Gerlach deflection it is 
only first derivatives of the
field that enter. In this sense the effective electric field 
components for the inverse square law electric field 
are 
\begin{align}
E^2 &= -2E_1\,\frac{x^2}{a_1},\quad E^2_{,2}=\frac{-2E_1}{a_1},\notag\\
E^3 &=   E_1\,\frac{x^3}{a_1},\quad E^3_{,3}=\frac{  E_1}{a_1}.
\label{eq:Bohr.3p}
\end{align}
As in a circular storage ring, in 
a more or less circular atomic orbit, treated classically, the spin will
precess around the normal to the orbit plane, both because of 
Thomas and S-G precession (which explicitly violates approximations
made up to this point.) Barring possible resonance
effects (a possibility we ignore, for now, but will revisit) 
any effective radial force
due to the Stern-Gerlach force acting on spin polarization components
lying in the orbit plane will necessarily average to zero. It is true,
however, that the spin component normal to the orbit plane, 
in our notation $\pm\hat{s}^{*3}$, will be conserved in this precession. 
Any radial force, in our notation $F^2$, will be constant, causing 
the radius of a circular orbit to depend (very weakly) on whether 
the spin is up or down.

In Figure~\ref{fig:E-B-Quadrupoles}
one notes that the closest match to this field pattern is
labeled ``erect electric quad''.  In this figure, with increasing
$x^2$ the electric field component $E^2$ varies proportionally with 
$x^2$, and similarly for $x^3$ and $E^3$.  
Comparing Eqs.~(\ref{eq:Bohr.3p}) with Eqs.~(\ref{eq:relSG.6Ee}), 
one also sees that the linearized Coulomb field components resemble 
those of an erect electric quadrupole. We can therefore apply
Eq.~(\ref{eq:relSG.5E}), with two of the electric field components
set to zero. The operative equations are
\begin{equation}
\begin{pmatrix} dp^2/dt \\ dp^3/dt \end{pmatrix}
  = 
\mu_e^*\,
\begin{pmatrix} 
 0   &  \hat s^{*3}E^3_{,3}/c \\
 0   & -\hat s^{*2}E^2_{,2}/c 
\end{pmatrix}
\begin{pmatrix} 1 \\ v/c \end{pmatrix}.
\label{eq:Bohr.4}
\end{equation}
We have already argued that any effect due to $\hat s^{*2}$
averages to zero, so we are left with a single equation
for the Stern-Gerlach force;
\begin{equation}
F^{SG}_{\pm}
 =
\frac{dp^2}{dt} 
  = 
\frac{\mu_e^*\hat s^{*2}v}{c^2}\,E^3_{,3} 
  = 
 \pm\frac{\mu_e^*v}{c^2}\,\frac{E_1}{a_1}.
\label{eq:Bohr.5}
\end{equation}
In the final step, as well as using
Eqs.~(\ref{eq:Bohr.3p}), the formula has been simplified
to cover just the cases of spin up and spin down.
The formula shows that the S-G force simply adds to
or subtracts from the Coulomb force, depending on whether
the spin is up or down. The ratio of S-G force to Coulomb
force is 
\begin{equation}
\frac{F^{SG}_{\pm}}{F^{\rm Coul}}
  =
 \pm\frac{\mu_e^*/c}{e}\,\frac{1}{a_1}\,\frac{v}{c} \\
  =
\pm\frac{1/2}{137.0359}\,\frac{v}{c} \\
  = \pm\frac{\alpha^2}{2}. 
\label{eq:Bohr.6}
\end{equation}
where the intermediate numerical ``coincidence'' is noted in passing.
Of course it cannot be a coincidence at all, since the problem being
addressed is the same as the problem Sommerfeld was attacking when
he defined his fine structure constant $\alpha$. 
It remains, however, to determine whether the magnetic moment value 
$\mu_B$=$5.788382\times10^{-5}\,{\rm eV/T}$ initially assumed, 
leads to sensible energy difference between
spin-up and spin-down energy levels in our toy
semi-classical model.

\subsection{Spin-Orbit Energy Shift} 
(Ignoring the reduced mass correction)
for electron motion in a circle of radius $a_1$ at electric field $E_1$,
the non-relativistic kinetic energy $K_{1,NR}$ and the 
the relativistic momentum $p_1=m\gamma v$ are  given by
\begin{align}
m_e\gamma\frac{v^2}{a_1}=eE_1, & \longrightarrow p_1c = \frac{eE_1a_1}{v_1/c} \notag\\
                              & \longrightarrow K_{1,NR} = eE_1a_1/2.
\label{eq:Orbit.1}
\end{align}
Should a next approximation be required for kinetic energy it can be 
obtained from
\begin{equation}
K_1 = \frac{eE_1a_1}{2}\,\Big(1 + \frac{3}{2}\,\frac{eE_1a_1/2}{m_ec^2}\Big). 
\label{eq:Orbit.2}
\end{equation}
At radius $a_1$ the potential energy and non-relativistic total 
energy are
\begin{equation}
V_1 =  -\frac{e^2}{4\pi\epsilon_0 a_1},\quad
\mathcal{E}_1 = -\frac{e^2}{8\pi\epsilon_0 a_1}.
\label{eq:Orbit.3}
\end{equation}
The only spin-orbit effect encountered so far is equivalent to a change
in central force. In the Bohr model, for a given state, the 
angular momentum quantum condition requires the angular momentum
$p_1a_1$ to be preserved, which requires
\begin{equation}
\frac{\Delta p_1}{p_1} = -\frac{\Delta a_1}{a_1}.
\label{eq:Ratios.1}
\end{equation}
For a circular orbit the momentum satisfies 
$p=\sqrt{2m_eK}=\sqrt{m_eeE_1a_1}$, which requires
\begin{equation}
\frac{\Delta p_1}{p_1}
 = 
\frac{1}{2}\frac{\Delta E_1}{E_1} + \frac{1}{2}\frac{\Delta a_1}{a_1}
\label{eq:Ratios.2}
\end{equation}
Combining Eqs.(\ref{eq:Ratios.1}) and (\ref{eq:Ratios.2}) produces
\begin{equation}
\frac{\Delta a_1}{a_1} = -\frac{1}{3}\frac{\Delta E_1}{E_1}.
\label{eq:Ratios.3}
\end{equation}
For circular orbits the total energy $\mathcal{E}$ is half the potential energy
energy $V_1=e^2/(4\pi\epsilon a_1)$, which varies inversely with $a_1$. 
Combining formulas 
\begin{equation}
\frac{\Delta\mathcal{E}_1}{\mathcal{E}_1}
 =
-\frac{\Delta a_1}{a_1}
 =
\frac{1}{3}\frac{\Delta E_1}{E_1}
 =
\frac{1}{3}\frac{F^{SG}_{\pm}}{F^{\rm Coul}},
\label{eq:Ratios.4}
\end{equation}
where, in the last step, $F^{SG}_{\pm}/F^{\rm Coul}$ is treated as a fractional
change in the central electric field. Substitution from 
Eq.~(\ref{eq:Bohr.6}) produces
\begin{equation}
\frac{\Delta\mathcal{E}_1}{|\mathcal{E}_1|}
 =
\frac{1}{3}\,\frac{F^{SG}_{\pm}}{F^{\rm Coul}}
 =
\pm\frac{\alpha^2}{6}.
\label{eq:Ratios.5}
\end{equation}
As emphasized previously, being classical rather than quantum
mechanical, this has just been a toy calculation intended mainly
as a ``sanity check'' of the relativistic formulation. Toward this end, 
one can compare result~(\ref{eq:Ratios.5}) with a quantum mechanical
calculation of spin-orbit doublet splitting. An exact comparison
is impossible because there is not exact correlation between Bohr
energy levels and quantum mechanical levels.

A somewhat similar, up-to-date, quantum mechanical
spin-orbit doublet separation calculation
can be copied from Leighton\cite{Leighton}, assuming $Z$=1, $n=1$,
$l=1$
\begin{align}
(a)&\ \frac{\Delta\mathcal{E}_1}{|\mathcal{E}_1|}
 = 
\frac{\alpha^2}{(2l+1)(l+1)},\quad\hbox{for}\quad j=l+1/2,\label{eq:Doublet.1}\\
(b)&\ \frac{\Delta\mathcal{E'}_1}{|\mathcal{E}_1|}
 = 
\frac{-\alpha^2}{l(2l+1)},\quad\hbox{for}\quad j=l-1/2,\ l\ne0.
\label{eq:Doublet.1p}
\end{align}
The classical and quantum calculations differ only in the numerical 
factors, which are $+1/6,-1/6$ in the classical calculation and
$+1/6,-1/3$ in the quantum mechanical calculation. This 
level of agreement more than satisfies the intended sanity check of 
our relativistic formulation.

Repeating a previous comment, the closeness of a classical to a quantum
result should not be surprising; the calculation is more-or-less
equivalent to a Sommerfeld calculation that was more-or-less successful
in interpreting spectral doublets even before 1920.

What is perhaps surprising, though, is that a crude Newtonian picture of 
an atom as a tiny storage ring, though wrong, is not flagrantly wrong. Included 
in this picture is the treatment of the S-G force as a tiny spin-dependent
alteration of the central electric force. Niceties such as Thomas precession,
and other delicate spin precession effects, are irrelevant in this 
picture because they simply average to zero without having any influence 
at all on the stable energy levels.
In accelerator physics practice one would be inclined
to blame the failure of the classical physics model on 
incorrect treatment of resonant spin effects. 
It seems the quantum mechanical Hamiltonian angular momentum 
addition formalism magically accounts correctly for this behavior.

That an atom can be visualized as a tiny storage ring is just a
curiosity.  But there is real content to the concept of a storage
ring as a giant atom, especially if the beam is polarized and the
polarization is ``phase-locked''. An atom cannot have very many electrons,
but a storage ring can have $10^{10}$. Even though these electrons have spreads
in position, slope, energy, and spin orientation, there can be enough 
of them for the polarization to be fed-back and phase-locked to an 
externally-imposed frequency. With spin precession proportional to
magnetic moment, which is known to the precision implied by the 
values listed in Eq.~(\ref{eq:Intro.1}), and all other coordinates
phase-locked as well, the storage ring has become a ``trap'' with
parameters stabilized to exquisite precision. Then, much like nuclear
magnetic resonance, there is then the possibility of detecting exquisitely
small effects superimposed on the centroid motion. Effects too small
to be observed over a single revolution can be magnified by the millions,
or even billions, of beam revolutions possible in a storage ring.

This concludes the theoretical discussion. Experimental details
follow.

\section{Practical Observation of S-G Deflection}
A recent conference presentation\cite{RT-Spin2016}, from a
Cornell, Jefferson Lab, University of New Mexico working
group, explains how the CEBAF 123\,MeV injection line can serve as one 
big Stern-Gerlach polarimeter measuring the polarization state of 
the injected electron beam. No physical changes to the line are 
required and (though not optimal) beam position monitors (BPMs) 
already present in the line can be used to detect the S-G signals.
Most of the remainder of this paper is drawn from that report.

The historical Stern-Gerlach
apparatus used a uniform magnetic field (to orient the spins) with quadrupole 
magnetic field superimposed (to deflect opposite spins oppositely) and a 
neutral, somewhat mono-energetic, unpolarized, neutral atomic beam
of spin 1/2 particles. For highly-monochromatic, already-polarized beams 
produced by Jefferson Lab electron guns, the uniform magnetic field has 
become superfluous, and every quadrupole in the injection line produces 
polarization-dependent S-G deflection. The absence of constant magnetic
field on the unperturbed electron orbit has the further effect of 
guaranteeing zero electric field at the electron's instantaneous position 
in its own (unperturbed) rest frame. Unlike in the actual Stern-Gerlach 
configuration, any non-zero electric field in such a frame is proportional 
to (miniscule) Stern-Gerlach deflections, and can therefore be neglected 
to good approximation. 

Starting from neutral silver atoms, with approximate velocity 500\,m/s 
in the original experiment, for which the angular deflections were roughly 
$\Delta\theta^{Ag}\approx 0.005\,$radians,
we can estimate the Stern-Gerlach deflections of 6\,MeV electrons in the
quadrupoles of a modern-day accelerator. In both cases the transverse force
is due to the magnetic moment of a single electron. Magnets in the CEBAF beam 
line are much like the original (1923) Stern-Gerlach magnets, though the
original magnetic field gradient$\times$length product was several times 
larger\cite{Friedrich-Herschbach} than for typical quadrupoles 
in the CEBAF injection line\footnote{Some parameters for the original 
Stern-Gerlach experiment were: central field 0.1\,T, peak field 
gradient 100\,T/m, $T=1350^{\circ}$K,
magnet length 0.035\,m, distance from magnet center to detecting film 0.02\,m.}. 
But, for simplicity, we compare the deflections of a silver 
atom and a free electron in identical magnets. 

The transverse force 
is the same irrespective of whether the electron is free or a valence electron in 
the silver atom. An anticipated deflection of electrons with $\gamma_e=12$ 
can be estimated from formulas for the angular deflection, 
$\Delta p_{\perp}/p_z$, for the ratio of force durations, $v^e/v^{Ag}$, and for the
ratio of longitudinal momenta, $p^{Ag}/p^e$:
\begin{align}
\frac{\Delta p_{\perp}}{p_z}
  &= {\rm Force}\frac{\rm duration}{p_z}\notag\\
\frac{v^e}{v^{Ag}}   
  &= \frac{3\times10^8}{500} = 6\times10^5\notag\\
\frac{p^{Ag}}{p^e}           
  &= \frac{M^{Ag}}{m_e}\,\frac{\gamma^{Ag}}{\gamma_e}\,
     \frac{v^{Ag}}{v^e}\notag\\
  &\approx\frac{108\times2000 m_e}{m_e}\,\frac{1}{12}\,\frac{500}{3\times10^8}\,
    \approx 0.03.
\label{eq:S-G-compare}
\end{align}
In the same magnet we expect 6\,MeV electron deflections of order
\begin{equation}
\Delta\theta^{6MeV e} \approx 0.005\times\frac{0.03}{6\times10^5} 
  \approx 2.5\times10^{-10}\,{\rm radians}.
\notag
\end{equation}
With drift lengths in the injection line being of order 1 meter, and allowing for
the somewhat smaller field integrals, this suggests that 
the \"Angstrom, which is equal to $10^{-10}$\,m, is an appropriate unit 
for expressing 
the S-G betatron amplitudes to be expected.

For a dedicated reconfiguration of the beamline optics, the S-G deflection could
be enhanced substantially. But, to minimize operational interuption, 
we first assume no CEBAF beamline changes whatsoever, so that investigations 
can be almost entirely
parasitic. Because the expected amplitudes are so small, 
we also consider reducing 
the beam energy to 6\,MeV or less (from 123\,MeV) throughout the line, 
by detuning the
intermediate linac section, to increase the S-G effect.

Dual CEBAF electron beam guns produce superimposed 0.25\,GHz 
(bunch separation 4\,ns) electron beams for which the polarization 
states and the bunch phases can be adjusted independently. For example, the 
(linear) polarizations can be opposite and the bunch arrival times adjusted so 
that (once superimposed) the bunch spacings are 2\,ns and the bunch polarizations 
alternate between plus and minus. The effect 
of this beam preparation is to produce a bunch charge repetition frequency of 
0.5\,GHz different from the bunch polarization frequency of 0.25\,GHz. This
difference will make it possible to distinguish Stern-Gerlach-induced 
bunch deflections from spurious charge-induced excitations.

Transverse bunch displacements produce narrow band BPM signals 
proportional to the $f_r$ Fourier frequency components of transverse beam 
displacement. Because linac bunches are short there can be significant 
resonator response at all of the strong low order harmonics of the 
0.25\,GHz bunch polarization frequency. The proposed S-G responses are 
centered at odd harmonics, $f_r=0.25, 0.75, 1.25$\,GHz. The absence of
beam-induced detector response at these frequencies greatly 
improves the rejection of spurious ``background'' 
bunch displacement correlated with bunch charge. For further background 
rejection the polarization amplitudes are modulated at a low, sub-KHz 
frequency which shifts the S-G response to sidebands of the central
S-G frequencies.  

\section{Stern-Gerlach Deflection of a Relativistic Particle}
This section is largely repetitive of previous derivations,
but from a more elementary, and more practical point of view. It is
specialized toward ongoing investigations using a Jefferson lab
injection line.

We are primarily interested in the Stern-Gerlach deflection caused 
by the on-axis passage of a point particle with velocity $v{\bf\hat z}$ 
and rest frame, transversely-polarized magnetic dipole moment vector 
$\mu_x^*{\bf\hat x}$, through a DC quadrupole, of length $L_q$, 
that is stationary in the laboratory frame $K$.

It is valid to formulate the calculation with an impulsive 
approximation, in which the integrated momentum imparted
to a particle passing through a quadrupole is small enough to justify 
neglecting the spatial displacement occurring during the encounter
and keeping track of only the angular deflection.\footnote{
For anomalous electron angular momentum $G=0.00116$ the spin precession
angle occurring during angular deflection $\Delta\theta$ of
approximately $G_e\gamma\Delta\theta/(2\pi)$ is negligible. 
All spins are taken to be purely horizontal (in the $x$-direction)
in both the laboratory and the electron rest frame. Similarly
there is no longitudinal magnetic field in either frame. 

On-axis
in a magnetic quadrupole there is neither magnetic, nor electric
field in either the K or the K' frame. Once an electron is
displaced by the S-G force, there is non-zero magnetic field in the 
K frame, and hence non-zero electric field in the K' frame. 
Solution of the orbit equation with this electric force included
is given in the appendix.}
One also notes the particle speed is conserved because it is only a 
longitudinal component of force that can change the particle speed.
The Stern-Gerlach deflection in the electron's instantaneous rest frame can 
simply be copied from well-established non-relativistic 
formalism\cite{Porter-SG}; the transverse force is given by 
\begin{equation}
F'_x = \mu_x^*\frac{\partial B'_x}{\partial x'}.
\label{eq:relSG.1}
\end{equation}
Following notation of Conte\cite{Conte-Stern-Gerlach}, the rest frame 
magnetic moment is symbolized by $\mu^*$ to stress that it is specific to the
rest frame, irrespective of whatever reference frame is being discussed.
A transverse spin in the laboratory is (by definition) also transverse
in the particle rest frame. And, concerning the present calculation, 
there is no issue of ``Lorentz transformation of spins or magnetic moments'',
since the S-G deflection is to be calculated in a frame of reference
in which the electron remains non-relativistic. In this frame
formula~(\ref{eq:relSG.1}) has been thoroughly confirmed by experiment. 

As viewed in the $K'$ rest frame, the passing magnet is Lorentz-contracted 
to length $L'_q=L_q/\gamma$,
the time spent by the particle in the magnetic field region is 
$L'_q/v$, and the integrated, rest frame transverse momentum impulse is 
\begin{equation}
{\Delta p'}_x
 = 
F'_x\,\frac{L'_q}{v}
 = 
\frac{\mu_x^*}{v}\,\frac{\partial}{\partial x'}\,(B'_xL'_q).
\label{eq:relSG.2}
\end{equation}
To determine $B'_x$ the laboratory magnetic 
field $B_{x}$ needs to be Lorentz transformed to the moving 
frame $K'$. This
produces both an electric and a magnetic field, but it is only the magnetic
field that produces Stern-Gerlach acceleration in the particle's rest frame. 
The Lorentz transformation
yields\cite{Jackson} $B'_x=\gamma B_x$. 
We conclude that the product
$B_xL_q=B'_xL'_q$ is the same in laboratory 
and rest frames. Since the 
displacement $x=x'$ and the transverse momentum component
${\Delta p}_x={\Delta p}'_x$ 
are also invariant for Lorentz transformation along the $z$ axis, 
Eq.~(\ref{eq:relSG.2}) becomes
\begin{equation}
{\Delta p}^{SG}_x
 = 
F_x\,\frac{L_q}{v}
 = 
\frac{\mu_x^*}{v}\,L_q\,\frac{\partial B_x}{\partial x},
\label{eq:relSG.3}
\end{equation}
and similarly for ${\Delta p}^{SG}_{y}$. The ``SG'' superscripts
have been introduced to distinguish Stern-Gerlach deflections from
Lorentz force deflections.

The conclusion so far is that formula~(\ref{eq:relSG.3}), derived 
historically using
non-relativistic kinematics, is valid even for relativistic particle speed.
Of course, because $v$ cannot exceed $c$, the transverse force saturates as
the particle becomes relativistic. Since the particle momentum continues to
increase proportional to $\gamma$, the S-G angular deflection in a fixed 
excitation quadrupole field falls as $1/\gamma$.

The magnetic field components of an erect DC quadrupole 
are given by
\begin{equation}
B_x=ky,\  B_y=kx,
\quad\hbox{where}\ 
k=\frac{\partial B_x}{\partial y}=\frac{\partial B_y}{\partial x}.
\label{eq:relSG.4pp}
\end{equation}
The quadrupole field in the original Stern-Gerlach experiment would,
in modern accelerator terminology, be referred to as ``skew''. 
The strong quadrupoles in the CEBAF line under consideration are 
``erect''.

Treating a quadrupole of length $L_q$ as a thin lens, the Lorentz force 
on a point particle of mass $m$ and charge $e$ traveling with velocity 
$v{\bf\hat z}$ through the quadrupole imparts momentum
\begin{equation}
\Delta{\bf p}
 =
{\bf F}(x,y)\,\Delta t
 =
eL_qk(y{\bf\hat y} - x{\bf\hat x}).
\label{eq:relSG.4p}
\end{equation}
The relativistic longitudinal particle momentum of the particle is 
$p=\gamma m v$ and its (small, linearized) electromagnetic angular 
deflections are given by 
\begin{equation}
\Delta{\theta_x}{\bf\hat x} + \Delta{\theta_y}{\bf\hat y}
 =
\frac{\Delta{\bf p}}{p}
 =
q_xx{\bf\hat x} + q_yy{\bf\hat y},
\label{eq:relSG.4q}
\end{equation}
where inverse focal lengths $q_x=1/f_x$ and 
$q_y=1/f_y$ of the quadrupole satisfy
\begin{equation}
q_x = -\frac{eL_qk}{p} 
    = -\frac{L_qc\partial B_y/\partial x}{pc/e}
    = -q_y.
\label{eq:relSG.4r}
\end{equation}
Meanwhile, the Stern-Gerlach deflections are given by 
\begin{equation}
{\Delta\theta}_{x}^{SG}
 =
\frac{{\Delta p}^{SG}_{x}}{p}
 = 
\frac{\mu_{x}^*L_qk}{pv},
\label{eq:relSG.9}
\end{equation}
and similarly for $y$.
Comparing with Eqs.~(\ref{eq:relSG.4r}), one sees that the 
Stern-Gerlach deflection in a quadrupole is strictly 
proportional to the inverse focal lengths of the quadrupole;
\begin{equation}
\boxed{
{\Delta\theta}_{x}^{SG}
 =
-\frac{\mu_x^*}{ec\beta} q_x,
\quad\hbox{and}\quad
{\Delta\theta}_{y}^{SG}
 =
\frac{\mu_y^*}{ec\beta} q_y,}
\label{eq:relSG.10}
\end{equation}
These formulas
are boxed to emphasize their universal applicability to
all cases of polarized beams passing through quadrupoles.
For all practical (electron beam) cases $\beta\approx1$.

As mentioned previously, the S-G deflection at fixed magnet excitation 
is proportional to $1/\gamma$. Yet, superficially, deflection 
formulas~(\ref{eq:relSG.10}) show no \emph{explicit} dependence on
$\gamma$ (such as, for example, the denominator factor 12, 
in Eq.~(\ref{eq:S-G-compare})).
This is only because the angular deflections are expressed
in terms of quadrupole inverse focal lengths. For a given quadrupole
at fixed quadrupole excitation, the inverse focal length scales as $1/\gamma$.
Expressing the S-G deflection in terms of inverse focal lengths has 
the effect of ``hiding'' the $1/\gamma$ Stern-Gerlach deflection 
dependence, which comes from the beam stiffness.

$\mu^*_x$ and
$\mu^*_y$ differ from the Bohr magnetron $\mu_B$ only by $\sin\theta$
and $\cos\theta$ factors respectively
Numerically, for one particular CEBAF quadrupole,
Eq.~(\ref{eq:relSG.10}) yields Stern-Gerlach-induced,
Courant-Snyder betatron amplitude proportional to
\begin{equation}
\sqrt{\beta_x}\,{\Delta\theta}_{x}^{SG}
  =
-(1.93\times 10^{-13}\,{\rm m})\,\sqrt{\beta_x}\,q_x,
\label{eq:relSG.3q}
\end{equation}
and similarly for $y$.
The $\sqrt{\beta}$ factor has been included because
the transverse displacement $\Delta x_j$ at downstream location ``j''
caused by angular displacement $\Delta\theta_i$ at 
upstream location ``i'' is given (in either plane) by
\begin{equation}
\Delta_j = \sqrt{\beta_j\beta_i}\,q_i\sin(\psi_j-\psi_i).
\label{eq:relSG.3p}
\end{equation}
where $\psi_j-\psi_i$ is the betatron phase advance from ``i'' to ``j'',
and $\Delta_j$ stands for either $\Delta x_j$ or $\Delta y_j$.

\section{S-G Specific Beam Preparation}
The smallness of the S-G signal, especially relative to unintentional
charge-sensitive cavity responses, makes
it critical for the polarized beam to be prepared for maximum
rejection of spurious background. 

Recent ILC-motivated BPM performance 
investigations
\cite{InoueResonantBPM}\cite{WalstonBoogert-ILC-BPM}
are relevant to our proposed Stern-Gerlach (S-G) detection experiment.
Resonant beam position detection relies on two TM cavities. 
The charge-sensitive cavity (needed to normalize the charge)
is tuned to resonate in a transversely 
symmetric mode at the bunch frequency. The position-sensitive cavity
is tuned to resonate in a transversely anti-symmetric mode
at the same bunch frequency.

(By the Heisenberg uncertainty principle) it would not be feasible to locate 
a single mono-energetic electron with usefully small transverse accuracy. This 
makes the electron charge $e$ unnaturally small for present purposes. For
comparison we define a ``standard macro-charge'' as the charge of 
$N_e=10^{10}$ electrons,
which is a typical number of electrons in each bunch in an ILC BPM 
prototype test, for example at the KEK Accelerator Test Facility (ATF). 
Classical (rather than quantum) mechanics is adequate for treating the 
centroid motion of such a large number of electrons, even as regards their
mean spin orientation.

A CEBAF beam is CW, with beam current of, say, 160\,$\mu$A, which corresponds
to a current of about $10^5$ (just-defined) macro-charges per second. 
For S-G detection the \AA ngstrom is a convenient 
transverse length unit for S-G detection. For comparison, 
successful ILC operation the 
transverse beam positions need to be controlled to about $\pm$10\,\AA.

The bunch structures of the CEBAF injector (123\,MeV, 160\,$\mu$A, $0.5$\,GHz)  
and ATF (1.3\,Gev, 
$N_ee=10^{10}e$ macro-charge at $5$\,Hz pulse rate) are very different. 
We ignore the energy difference, which is thought to be unimportant for 
the comparison. For a typical cavity resonator quality factor of $Q_r=10^4$ and 
frequency of 1\,GHz, the cavity discharging time is far shorter
than the ATF repetition period. This makes it appropriate to treat the ATF
resonant response on a pulse-by-pulse basis. Essentially different in time 
structure, the CEBAF resonator response is continuous wave (CW) with the 
previously-defined macro-charges passing through the cavity at 100\,kHz rate. 

\begin{figure}[h]
\centering
\includegraphics[scale=0.18]{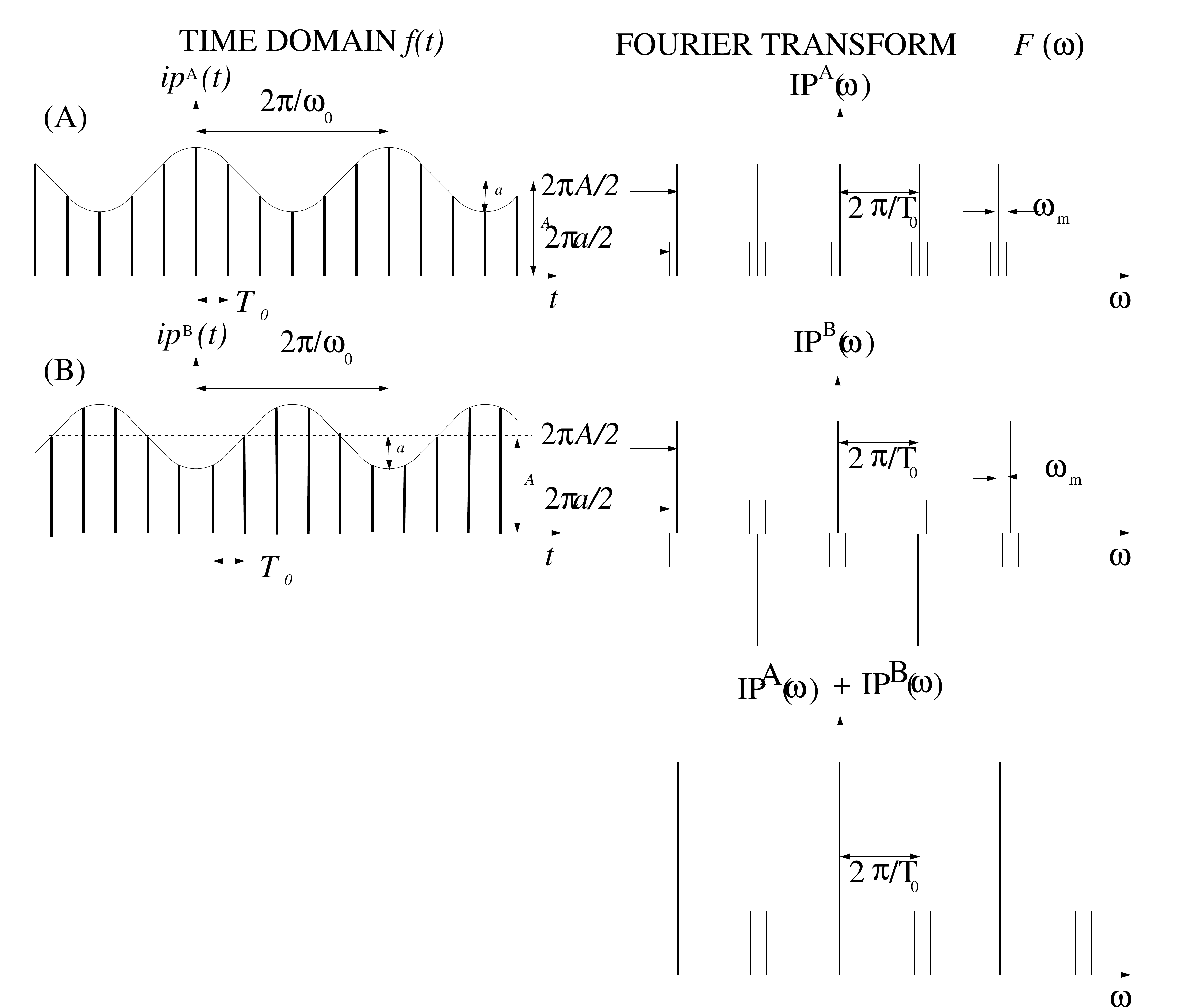}
\vskip -0.2cm
\caption{\label{fig:FourierTransforms}Time domain and frequency
domain beam pulses for the A and B staggered, modulated-polarization beams. 
It is current-weighted polarization spectra that are plotted
in these figures. The current spectra themselves are obtained by suppressing
the modulation sidebands. In the A+B spectra the odd harmonics of
beam current cancel, effectively doubling the fundamental current frequency 
from 0.25\,GHz to 0.5\,GHz. But the current-weighted polarization side bands 
survive as odd harmonics of  0.25\,GHz.
}
\end{figure}

In a linac beam line, the fact that each bunch passes an
S-G sensitive BPM only once, makes it hard to arrange for the polarization of 
successive bunches to be different. But, as already explained, high frequency 
bunch polarization modulation frequency is made possible
by superposing staggered bunch trains having different polarizations.
Figure~\ref{fig:FourierTransforms} illustrates our planned, 
superimposed CEBAF bunch
train. Bunches are labeled A in one of two pre-superimposed bunch trains
and B in the other. Time domain plots are on the left, frequency spectra
on the right. The foreground S-G betatron signal oscillates at 
(harmonics of) 0.25\,GHz, while the background charge signal oscillates at 
(harmonics of) 0.5\,GHz. For resonant cavity BPMs the S-G detector
would be tuned to a harmonic of the 0.25\,GHz fundamental, for example to
the third or fifth harmonic, for more convenient (smaller) cavity dimensions.

We assume the polarization of the superimposed A and B beams are also modulated
with (low) frequency $\omega_m$. The time domain, $i\,p(t)$ current-polarization
products of the separate A and B beams are then given by
\begin{align}
i\,p^A(t) &= \sum_{n=-\infty}^{\infty}\delta(t-nT_0)(A+a\cos\omega_mt)
\label{eq:Fourier.1}\\
i\,p^B(t) &= \sum_{n=-\infty}^{\infty}\delta(t-T_0/2-nT_0)(A-a\cos\omega_mt).
\notag
\end{align}
and are plotted on the left in Figure~\ref{fig:FourierTransforms}. 
The modulation amplitude $a$ is drawn much smaller in magnitude
than the un-modulated polarization amplitude $A$. But over-modulation, 
with values of $a$ as great as $2A$, to maximize the side-band amplitudes, 
might be practical.
There are two essential differences between the A and B beams. 
The beam pulses are shifted in time by one half cycle and the sign 
of the modulation is reversed.
The modulation frequency $\omega_m$, for which the frequency is expected to
be about 1\,kHz, is exaggerated by many orders of
magnitude in this figure, since $f_0=1/T_0$ is about 0.75\,GHz.
Champeney\cite{Champeney} gives the A-beam, cosine-modulated 
current-polarization Fourier transform $IP^A(\omega)$ to be
\begin{align}
&I\,P^A(\omega)
 =
\sum_{n=-\infty}^{\infty}\,
\frac{2\pi}{T_0}\,
\bigg(
A\delta\Big(\omega - n\frac{2\pi}{T_0}\,\Big)
 +  \label{eq:Fourier.2}\\
&+\frac{a}{2}\,\delta\Big(\omega - n\,\frac{2\pi}{T_0} + \omega_m\Big)
 + \frac{a}{2}\,\delta\Big(\omega - n\,\frac{2\pi}{T_0} - \omega_m\Big)
\bigg). 
\notag
\end{align}
The Fourier transform of the B-beam, sine-modulated, 
current-polarization Fourier transform is obtained by multiplying
by the time-shift factor, $e^{-iT_0\omega/2}$ for which, when it is moved 
inside the summation, its $\omega$ factor can be replaced
by $2\pi n/T_0$, due to the delta function having
argument $\omega - 2\pi n/T_0$. The resulting $(-1)^n$ factor causes the sign 
alternation exhibited in the middle right graph in 
Figure~\ref{fig:FourierTransforms}.
Because the modulation frequency is
so low the corresponding time shift of the modulation is being neglected.

\section{Signal Levels and Background Rejection}
According to Eqs.~(\ref{eq:relSG.10}) the transverse displacement
magnitude $\Delta$ at distance $L$ downstream of a quadrupole of strength $q$ 
is given by
\begin{align}
\Delta &= (1.932\times 10^{-13}\,{\rm m})qL \label{eq:SignalLevels.1}\\
   &\overset{\rm e.g.}{\ \approx} (2\times 10^{-13}\,{\rm m}) 
                         \times{\rm 1/m}\times{\rm 10 m}
 = 0.02\times10^{-12}\,{\rm m}.
\notag
\end{align}

The installed CEBAF beam position monitors are ``antenna BPM's'', each
consisting of four short (approximately 4\,cm long), longitudinal, probe antennas,
symmetrically-located azimuthally, within cylindrical beam tubes.
Not themselves being narrow band, these BPMs are more noise-sensitive than 
resonant BPMs. But they have the 
important advantage of responding to both symmetric and 
anti-symmetric modes over large frequency range, for example at both 0.75\,GHz 
and 1.0\,GHz, with the responses easily separable by narrow band external 
spectrum analysis.

Barry\cite{CEBAFbpms} gives the transverse impedance of standard CEBAF BPMs to
be $Z_{\perp}=3800$\,V/m. Taking 200\,$\mu$A as a satisfactory beam current, the 
corresponding BPM power level would be 
\begin{align}
P &\approx (0.0002)^2\times 3800\times(0.02\times10^{-10}) \notag\\
     &= 3.0\times10^{-16}\,{\rm W} = -125\,{\rm dBm}.
\label{eq:SignalLevels.2}
\end{align}
At room temperature the thermal noise in a 1\,Hz bandwidth is given by
\begin{align}
P_{\rm noise} &= kT\Delta f
 = (1.38\times10^{-23}{\rm J/K})\times(293\,{\rm K}\times 1\,{\rm Hz}\notag\\
 &= 4.1\times10^{-21}\,{\rm W}
 = -174\,{\rm dBm}
\label{eq:SignalLevels.3}
\end{align}
This calculation suggests that, with seconds-long data collection, 
the S-G signal level will be large enough to be distinguishable from 
thermal noise. (A much lower thermal noise floor could be achieved 
with cryogenic detection.)

A more serious impediment to S-G detection is spurious cavity response
to bunch charge rather than to bunch polarization. We now review the
procedures to be employed in distinguishing S-G signals from background.

\noindent
{\bf Centered cavity. }
Conventional BPM beam centering relies on exact cavity centering for which,
ideally, there is no direct charge excitation of the position sensitive 
cavity.  Roughly speaking, the previously-described 
ILC BPM prototypes have so far achieved absolute 
transverse position reproducibility of $\pm15$\,nm, for bunch to bunch variation 
of beam bunches containing $N_e=10^{10}$ electrons. This is roughly an order 
of magnitude greater than (i.e. inferior to) their theoretical-minimum expected 
resolution of $\pm1.8$\,nm. The authors (persuasively) ascribe their BPM 
performance short-fall primarily to error sources other than thermal noise, 
such as instrument imperfections or cross-talk from spurious, mode-forbidden 
response to bunch charge. 

The ``good news'' to be drawn from the ILC $\pm1.8$\,nm noise floor is that,
with long time averaging, because of the high average CEBAF beam current,
coherent betatron oscillation amplitudes as small 
as, say, 0.02\,\AA, can be expected to emerge from thermal noise, 
even with room 
temperature detectors. The ``bad news'' is that there is little 
reason to suppose that cavity-centering S-G selectivity 
(relative to spurious background excitation) can be improved appreciably 
by increasing data collection times. Based on this estimate,
an S-G induced betatron amplitude of 0.02\,\AA, though distinguishable from 
thermal noise in a single, carefully-centered, conventional resonant BPM, can 
be expected to be dwarfed by a background/foreground ratio of more than
one thousand. This limitation is specific to the beam position and beam charge 
signals occurring at the same frequency, as in conventional beam position 
centering.

Based in this discuaaion,
accuracies as small as $\pm$20\,\AA\ should be achievable with centered 
transverse resonant BPMs. We are striving to measure betatron amplitudes
1000 times smaller. Mainly we need to make the case that rejection of spurious 
BPM signals caused by the beam charge (rather than the beam polarization) can 
be improved by three orders of magnitude compared to their influence on currently 
achievable transverse resonant BPM accuracy. The further selectivity improvement 
factors to be expected are surveyed next. 

\noindent
{\bf Disjoint polarization and charge frequencies. }
As explained earlier,
the polarized beam will be tailored so that the bunch polarization and 
bunch charge frequencies are different. In this condition
the BPM cavity is sensitive to polarization at one frequency (0.75\,GHz) 
and to charge at a different frequency (such as 0.5, or 1.0\,GHz). Ideally,
the resulting frequency domain filtering will suppress the spurious
background response by many orders of magnitude. More realistically,
there will still be background response, for example due to the small
Fourier component of charge excitation due to not-quite-cancelling
beam A and beam B currents. 

Empirical beam steering to null ``common mode'' BPM responses at both even
and odd harmonics of 0.25\,MHz (which would all vanish for 
ideal beam preparation) is especially useful for rejecting spurious background
excitation. This cancels both off-axis 
background excitation at the fundamental beam charge frequency and charge-imbalance 
background ``leakage'' from even harmonics to odd harmonics, while preserving the 
foreground S-G response differentially in the odd harmonics.

One can expect significant background/foreground suppression from these
common mode suppression and differential mode frequency domain filtering 
measures---perhaps three orders of magnitude. 

\noindent
{\bf Sideband shift of polarization frequency. }
As explained previously, the effect of low frequency modulation of the beam 
polarizations is to shift the S-G response to sidebands of the central
cavity resonance. To the extent the beam currents are unaffected by
this modulation, the sideband response will provide a pure S-G signal.
In practice the beam currents will, in fact, also be weakly modulated
which will allow some background signal to leak out to the side-band
frequencies. Still one can expect significant 
background/foreground suppression---perhaps two orders of magnitude.

\noindent
{\bf Multi-detector response modeling. }
In the CEBAF line under discussion the foreground S-G response 
will be monitored, with various (well known) degrees of sensitivity, in both 
$x$ and $y$ planes, at 19 BPM locations. The extent to which the
beam charges are (unintentionally) being low-frequency modulated at the gun 
can be parameterized with a few parameters, say 4, the main one describing
charge imbalance. Modulation of initial (low energy) beam angles will also
mimic S-G signals in individual BPMs. 
The corresponding betatron amplitudes are adiabatically 
damped by the subsequent acceleration, but they may remain significant.
But there is no reason to suppose that the downstream sensitivity to
starting beam conditions is correlated with S-G sensitivity. If true,
any spurious side-band responses can be subtracted by a model fitted
to match the total responses at all BPMs. Perhaps two orders of magnitude
selectivity improvement can be achieved.

\noindent
{\bf Lock-in signal detection. }
Though not mentioned previously, it is also true that the resonator
responses will be coherent with the beam bunch frequency. By lock-in
detection, the in-phase and out-of-phase S-G sideband deflections
can be determined individually. As well as improving noise rejection,
this can serve to corroborate the response model just described. 
Perhaps one more order of magnitude selectivity improvement can be 
achieved.

Multiplied together, the possibility of achieving eight orders of
magnitude rejection of spurious background has been described. This 
seems conservatively greater than the required three orders of magnitude
indicated earlier. Another factor of 6 improvement might be achieved by
lowering the beam energy entering the transfer line from 6\,MeV to 
1\,MeV. This would be satisfactory for an initial proof of principle,
but would not be tolerable for eventual routine polarimetry during
production CEBAF running.

\section{Recapitulation and Acknowledgements}

The intended purpose for this paper has been to provide technical back-up
for an experimental program confirming our understanding of 
Stern-Gerlach deflection. Apart from the fundamental physical significance
of Stern-Gerlach physics, the use of the S-G effect for non-destructive,
high analysing power, relativistic electron polarimetry, is an essential 
next step in the gradual improvement of spin control in storage rings. Such
polarimetry will be obligatory for any storage ring determination of the
electric dipole moment of the electron.

Discussion in this paper of practical S-G detection has been limited to
linac electron beams, because this is the only equipment that is 
immediately available 
without substantial development effort. The eventual application of 
Stern-Gerlach deflection for beam polarimetry is needed far more for
polarization control in storage rings, especially for ``frozen spin''
operation. In a linac a single electron passes a single BPM only once;
in a storage ring each electron passes through the same BPM 
millions of times per second. This greatly favors a storage ring over
a linac for S-G detection. With proper phase control, in frozen spin,
or pseudo-frozen spin storage ring operation, the repetitive passage through
the same high-Q cavity can, in principle, increase the S-G response
by another huge factor. Furthermore the beam current in storage ring,
even with polarized beam, can be far higher in a storage ring than in
a linac. These considerations make Stern-Gerlach polarimetry in a
storage ring very promising.

As mentioned previously, the experimental methods
described in the previous section have been developed 
by a working group made up of the
authors of reference~\cite{RT-Spin2016}. These ideas depend on experience 
gained during the design and implementation of polarized electron beams 
for the CEBAF accelerator at Jefferson Lab. Credit for the most important 
idea of all (toggling polarization) which is expected to make S-G 
detection possible, belongs to the Jefferson lab injection
group. This capability depends critically on polarized beam preparation 
tools developed for the Jefferson lab physics program.  

I would like to acknowledge the numerous conversations I have had on
this theoretical aspects of this subject with Saul Teukolksy and Eanna 
Flanagan. As already explained, experimental aspects have been developed 
in collaboration with Joe Grames, Reza Kazimi, Matt Poelker, Riad Suleiman,
and Brock Roberts.

\section{Appendix}
S-G deflection of an electron passing on-axis through a quadrupole
has been calculated approximately in the body of the paper. For purposes
of planning experimental corroboration of the formulas this degree of
accuracy is sufficient. Deviations from the simple calculation are
considered in this appendix.
The magnetic and electric fields in the electron 
rest frame are 
\begin{equation}
{\bf B'} = \gamma         k ( y'{\bf\hat x} + x'{\bf\hat y}) ,\quad 
{\bf E'} = \gamma \beta c k (-x'{\bf\hat x} + y'{\bf\hat x}).
\label{eq:appendix.1}
\end{equation}
For a vertically polarized electron the only rest frame magnetic
moment is $\mu_y^*$, the only non-zero S-G force is horizontal
\begin{equation}
{F^{SG}}'_x
 = \frac{\partial(\pmb\mu\cdot{\bf B})}{\partial x'}
 = \mu_y^*\,\frac{\partial B'_y}{\partial x'}
 = \mu_y^*\gamma k,
\label{eq:appendix.1p}
\end{equation}
and, for an on-axis particle, this is the only force. To lowest
approximation the motion in the rest frame is purely horizontal.
To next approximation, as $x$ deviates from zero, there is 
electromagnetic force
\begin{equation}
{F^{EM}}'_x
 =
-eE'_x - e\beta c \frac{dx'}{dt'}\,{\bf\hat x}\times B'_y{{\bf\hat y}}\Big|_x
 =
e \gamma \beta c k x'.
\label{eq:appendix.1q}
\end{equation}
The horizontal equation of motion is
\begin{equation}
\frac{d^2F'}{d{t'}^2}
 = 
\frac{e\gamma\beta ck}{m_e}\,\Big(x' + \frac{\mu_x^*}{e\beta c}\Big)
\equiv
{\omega'}^2\tilde{x'},
\label{eq:appendix.2}
\end{equation}
where $\tilde{x'}=x' + \mu_x^*/(e\beta c)$. Matching initial
conditions, the solution, in ascending powers of $\mu_x^*$, is
\begin{equation}
\tilde{x'}
 = \frac{\mu_x^*}{e\beta c}\,\cosh{\omega' t'}
 = \frac{\mu_x^*}{e\beta c}\,\Big(1+\frac{(\omega't')^2}{2} + \cdots\Big).
\label{eq:appendix.3}
\end{equation}
Using $x'$=$x$, after rest frame time duration $t'=(L_q/\gamma)/v$, 
the exit laboratory particle displacements and slopes are
\begin{equation}
x_{\rm exit} = \frac{\mu_x^*k L_q^2}{2m_e\gamma v^2},\quad
\Delta\theta^{SG} \approx \frac{\mu_x^*k L_q}{m_e\gamma v^2}.
\label{eq:appendix.4}
\end{equation}
This agrees with Eq.~(\ref{eq:relSG.9}).

\end{document}